\documentclass[aps,prl,nofootinbib,showpacs,twocolumn]{revtex4-1}
\usepackage{graphicx}
\usepackage{dcolumn}
\usepackage{amsmath}
\usepackage{latexsym}
\usepackage{mathrsfs}

\newcommand{\be}{\begin{equation}} 
\newcommand{\ee}{\end{equation}} 
\newcommand{\bea}{\begin{eqnarray}} 
\newcommand{\eea}{\end{eqnarray}} 

\begin{document}

\title{ Link-weight distribution  of microRNA co-target networks\\ exhibit universality }
\author{Mahashweta Basu$^1$, Nitai P. Bhattacharyya$^2$, P. K. Mohanty$^1$}
% \email[E-mail address: ]{mahashweta.basu@saha.ac.in}
\affiliation {$^1$Theoretical Condensed Matter Physics Division,
$^2$Crystallography and Molecular Biology Division,\\ Saha Institute of Nuclear Physics,
1/AF Bidhan Nagar, Kolkata, 700064 India.}

\begin{abstract}
MicroRNAs (miRNAs) are   small non-coding  RNAs   which  
regulate  gene expression by  binding to the $3'$ UTR of the 
corresponding messenger RNAs.  We construct  miRNA co-target  networks for 
$22$ different species  using a target prediction database,  MicroCosm Tagets. 
The miRNA pairs of  individual species having  one or more common target genes  
are connected  and the number of co-targets   are    assigned 
as the weight  of   these links. We show that the link-weight  distributions 
of  all the species collapse remarkably onto each other 
when scaled suitably. It turns out  that the  scale-factor  
is a measure of complexity   of the species. A  simple model,  where  targets are chosen 
randomly by miRNAs, could provide the correct scaling function   and 
explain the universality. 

\end{abstract}

\maketitle

\section{Introduction} 
Biological functions  occur in living cells  through bio-chemical interactions  of 
proteins. It is  a central dogma \cite{cdogma} of molecular biology that 
protein synthesis occurs inside the cell in two steps, (i) {\it transcriptions}, where  information from 
genes are  transfered to the messenger RNA  (mRNA) and (ii) {\it translation}, where 
information coded in mRNA is translated into specific  sequence of amino acids (proteins). 
The protein  densities in the  cell are primarily   regulated by  transcription 
factors \cite{TF}, however  recent studies 
\cite{book,farh} show that  a set of  small non-coding 
single stranded RNAs, namely micro-RNAs (miRNAs), also   act as 
secondary regulators.  MicroRNAs are produced from either their own genes 
or from introns. MicroRNAs are about $20$ nucleotides long,  
they  usually bind  to the $3'$ UTR  of the mRNA  inhibiting  their functionality.
Several computational tools \cite{majoros,grimson} have been developed to identify, 
firstly  the genomic sequences which can transcribe miRNAs  and their  possible   targets.   
It  has been estimated  that  \textit{Homo sapiens}  have  $851$  miRNAs \cite{miRBase} 
and their predicted targets constitute about $90\%$  of the  total  genes \cite{miranda}. 
Experimental validation of such predictions are, however, largely lacking.

Being a secondary  regulator,  miRNAs  usually  repress the   gene expression   marginally. 
Thus   it is natural to expect  that  cooperative  action   of miRNAs   are  needed   for 
alteration  of  any biological function or pathway.  
Recent studies \cite{xu} have revealed  this co-operativity using 
miRNA co-target networks,   constructed by   taking  miRNAs  as nodes 
connected by weighted links where the weight corresponds to the common targets of 
the  connecting pair. 
 Apparently $50\%$ miRNAs  in \textit{Homo sapiens} provide all essential regulations by forming several
small miRNA clusters \cite{mookherjee}. Study of   miRNA co-target networks   for different 
species  reveal that  these networks are   quite similar and  are robust  against 
random  deletion of nodes \cite{Lee}.

 \begin{figure}[h]
 \includegraphics[width=8.5cm,bb=14 14 776 277]{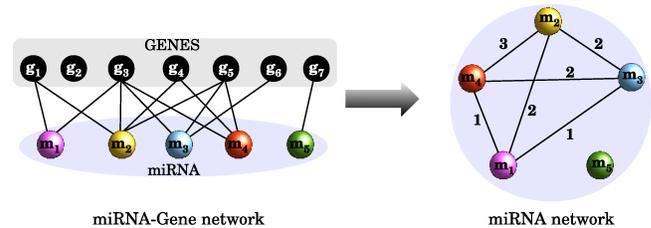}
\caption{ Schematic miRNA co-target network  of an example species 
having  $N=7$ genes and $M=5$ miRNAs. The  bipartite network (left) shows 
 genes, targeted  miRNAs.  The miRNA co-target network (right) is formed  by joining  
miRNA pairs which have at least one  common target. The  weights (number 
of co-targets)  are  written beside  each  link. }
 \label{fig:cartoon}
\end{figure}

In this article we show that the   distributions of weights  of  miRNA co-target networks 
are strikingly universal.  The universality can be explained  through a   simple   model  
that  assumes unbiased  binding   of  miRNAs    with  the available mRNAs.  Since species 
have different number of miRNA and genes, the mean and  standard deviations (SD)  of their
weight distributions  naturally differ. However, the  distributions are related to 
each other  through a simple scaling   indicating that  the underlying binding mechanism is unbiased.

\section{\bf MiRNA co-target network} 

To construct the miRNA  co-target network we use  a web resource 
MicroCosm Targets \cite{microcosm} which provides 
computationally predicted targets  of  microRNAs across many species. 
To predict the targets MicroCosm uses  the  miRNAs  sequences from a well known 
miRNA  prediction database miRBase \cite{miRBase}   and  genomic sequences  
from EnsEMBL \cite{EnsEMB}.  The number of predicted  miRNAs $M$ and  
the  total number of genes $N$ are listed in Table \ref{table:I}   
for  $22$ different species.  Note that the species 
considered   here  are quite sparse with respect to their class.
For all the species,   a miRNA  can  target  several  genes  and a gene   can also be targeted by several  miRNAs. 
This gives rise to the possibility that a pair of miRNA can have more than  one   common targets or co-targets.
The co-target network is constructed   separately  for  each species by  taking  
their miRNAs  as  nodes.  A miRNA  pair having   $w>0$ number of common target genes  are then 
connected    if by a link  of weight $w$.  The detailed procedure is described  
schematically in the fig. \ref{fig:cartoon}.

\begin{figure}[h]
 \includegraphics[width=8.5cm]{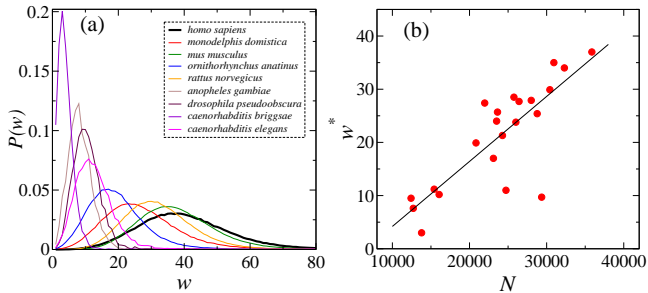}
 \caption{ (a) The link-weight  distributions  $P(w)$ for 
 some representative species. (b) The peak position $w^*$ depends linearly 
 on the number of genes $N;$   the best fitted line has slope  $=0.0012(1)$ 
 and $y$-intercept $=-8.021$.}
 \label{fig:Pw_Nw}
\end{figure}

 In these networks the weight  of the links, i.e.   the  number of co-targets  
 of a pair of miRNAs, vary  in a wide range.  
 For \textit{Homo sapiens}   the weights are bounded  in the range $1 \le w \le 1282$, whereas 
 it varies  in a smaller range  $1 \le w \le 513$ for \textit{C. Elegans}.  
 The distribution function $P(w)$ of the weights $w$ are  
 calculated separately for  $22$  species. Figure  \ref{fig:Pw_Nw}(a)   
 shows $P(w)$ vs. $w$  for some representative  species. 
 All these distribution functions  show a  single peak at some  value of 
 $w$, say $w^*$,  which is different  for different  species.  The  values  
 of $w^*$ are also listed in Table  \ref{table:I}.
 It is natural to expect a higher $w^*$   for  the species  which has larger number 
 of genes.  We find, to a reasonable approximation,  that $w^*$ varies 
 linearly with the  number of respective  genes  $N$  (see   fig. \ref{fig:Pw_Nw}(b)).

\begin{table*}
\begin{minipage}[b]{0.72\textwidth}
\hspace*{0cm}
\caption{List of species studied here.}
\label{table:I}
\begin{tabular}{|c|l|rrrr|rr|}
  
\hline
Sl.& {\bf Species} (cell types \cite{types}) & {\bf M} & {\bf N} & $w^*$ & $\lambda$ & $\bar n$ & $s$ \\
 \hline
1& \textit{Aedes aegypti} (-)  & 82 & 16059 & 10.2 &  0.32 & 388.71 & 47.94 \\
2& \textit{Anopheles gambiae} (65) & 82 & 12708 & 7.6 & 0.25 & 303.33 & 38.39 \\ 
3& \textit{Drosophila pseudoobscura} (-) & 88 & 12416 & 9.5 & 0.30 & 247.67 & 38.79 \\ 
4& \textit{Drosophila melanogaster} (63) & 93 & 15416 & 11.2 & 0.34 & 392.31	& 55.53 \\ 
5& \textit{Caenorhabditis briggsae} (35) & 135 & 13785 & 3.0 & 0.15 & 161.45 &	37.39 \\ 
6& \textit{Caenorhabditis elegans} (35) & 136 & 24728 & 11.0 & 0.40 & 513.04 & 84.90 \\ 
7& \textit{Gasterosteus aculeatus} (-) & 172 & 26423 & 27.7 & 0.65 & 824.63 & 111.07 \\ 
8& \textit{Oryzias latipes} (-) & 172 & 23514 & 24.0 & 0.59 & 729.27 & 98.42 \\ 
9& \textit{Takifugu rubripes} (120) & 173 & 21972 & 27.4 & 0.64 & 761.92 & 97.76 \\ 
10& \textit{Tetraodon nigroviridis} (120) & 174 &28005 & 27.9 & 0.63 & 828.94 & 100.69\\ 
11& \textit{Xenopus tropicalis} (130) & 199 & 24272 & 21.3 &  0.58 & 669.93 & 96.35 \\
12& \textit{Danio rerio} (120)  & 233 & 28744 & 25.4 & 0.62 & 792.40 & 103.90\\
13& \textit{Monodelphis domestica} (-) & 644 & 26013 & 23.8 & 0.79 & 799.11 & 150.23\\
14& \textit{Gallus gallus} (152)  & 651 & 20842 & 19.9 & 0.58 & 608.95 & 86.90\\
15& \textit{Macaca mulatta} (-)  & 656 & 32302 & 34.0 & 0.88 & 954.25 & 141.03\\
16& \textit{Pan troglodytes} (175)  & 662 & 29355 & 9.7 & 0.32 & 227.16 & 32.08\\
17& \textit{Canis familiaris} (160)  & 668 & 23628 & 25.7 & 0.78 & 768.99 & 124.87\\
18& \textit{Ornithorhynchus anatinus} (-)  & 668 & 23097 & 17.0 & 0.59 & 624.54 & 117.07\\
19& \textit{Bos taurus} (-)  & 676 & 25759 & 28.5 & 0.82 & 814.99 & 126.36\\
20& \textit{Rattus norvegicus} (160)  & 698 & 30421 & 29.9 & 0.75 & 891.74 & 131.47\\
21& \textit{Mus musculus} (160)  & 793 & 30484 & 35.0 & 0.84 & 885.63 & 127.49\\
22& \textit{Homo sapiens} (175) & 851 & 35864 & 37.0 &  1.00 & 959.03 & 147.03\\
\hline
\end{tabular}
\label{tab:singlebest}
\end{minipage}
\begin{minipage}[b]{0.18\textwidth}
\hspace*{-3 cm}\begin{tabular}{c}\centering
\includegraphics[height=8.5cm,bb=14 14 214 736]{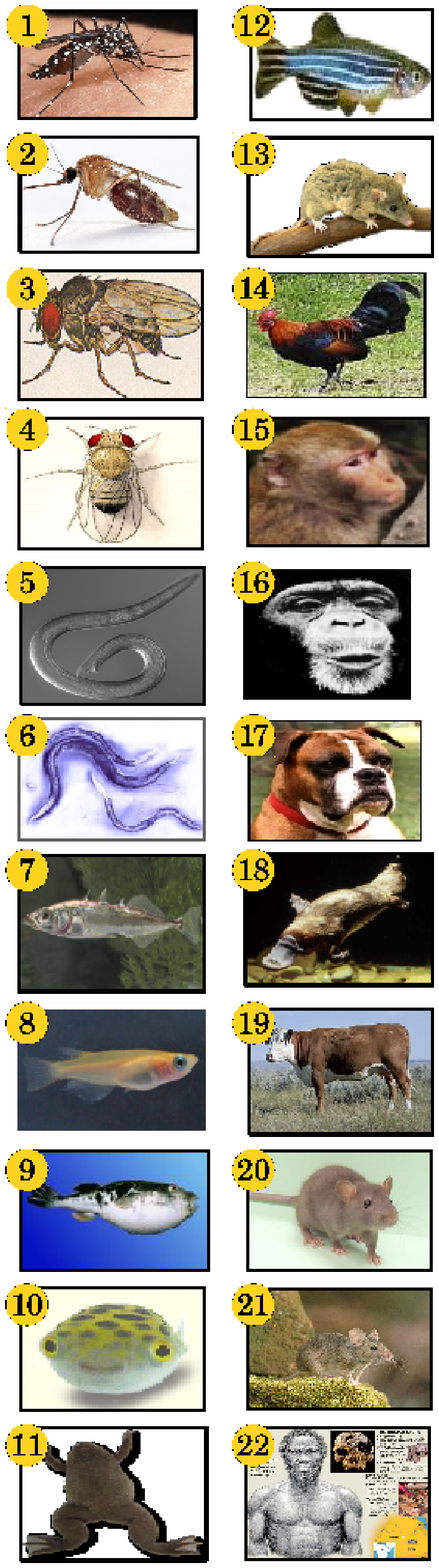}
\end{tabular}
\label{tab:twobest}
\end{minipage}
\end{table*}

\section{Universality}
\begin{figure}[t]
 \centering
 \includegraphics[width=8.6cm]{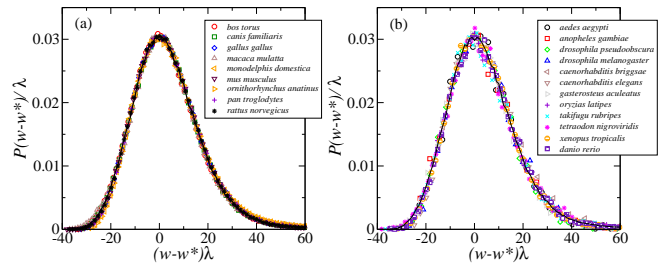}
\caption{$P(w)$ for (a) species with $(M > 300)$ and (b) the rest, are collapsed onto the
distribution curve for \textit{Homo sapiens} (solid line). Note that in (b) the fluctuation is larger
as these species has lower number of miRNAs.
}
\label{fig:collapse}
\end{figure}

 The distribution  functions  $P(w)$s  show an interesting scaling behaviour, $i.e.$ they 
could be collapsed onto a unique  scaling  function,  even though a large diversity is 
present among  the species. To observe the collapse, the distribution functions are first 
shifted  using a linear 
 transformation  $w \rightarrow w-w^*$ which  bring  the peaks of $P(w)$ to  origin  and 
 then   both the axises  are re-scaled  suitably using a scaling parameter  $\lambda$.  The
probability density   function (PDF) 
obeys  a scaling relation   $P( \lambda w)=  P(w)/\lambda$   to assure the normalization 
$\int_0^\infty  P(w) dw =1.$   Thus  a linear  shift and a re-scaling,   done here,  does not alter the 
functional form.   
In fig. \ref{fig:collapse}(a)  we have plotted 
$P(w-w^*)/\lambda$ vs. $ (w-w^*)\lambda$  for species  having  larger number 
of miRNAs $M> 300,$  where $\lambda$ is chosen such that  the shifted distribution 
functions are collapsed best  on to  the unscaled  data  of one of the species 
(here  \textit{Homo sapiens}). Data-collapse for  species  having  lesser number 
of miRNAs $M<300$  are shown separately in  fig. \ref{fig:collapse}(b)  as  they
have  large fluctuations which   obstruct the visual clarity.   Clearly, the  
rescaled $P(w)$ in both figures  matches remarkably  with the PDF  for 
\textit{Homo sapiens} $P_h(w)$  (shown  as a thick solid line).
This suggests  that  a universal functional form    governs the   distribution of 
number of co-targets    across  a wide class of species,  
even though, the miRNAs and their predicted targets are  quite  different among  species.

\begin{figure}[h]
 \centering
 \includegraphics[width=4.2cm]{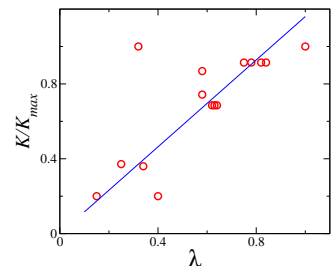}
 % complexity.eps: 0x0 pixel, 300dpi, 0.00x0.00 cm, bb=
\vspace*{-.4cm}\caption{ The scale factor $\lambda$   is proportional  the number of 
cell types $K$ (here normalized by  $K_{max}=175).$  The proportionality constant is $1.16(8).$}
 \label{fig: complexity}
\end{figure}

At this point the following comment is in order :  the  scale-factor  $\lambda$  (see Table \ref{table:I}) used for collapsing   $P(w)$s   
of   different species can be considered as a measure  of   morphological complexity  in animal  evolution.    In   fig. \ref{fig: complexity} we plot 
$\lambda$   versus the   number  of cell  types $K$  of  the respective species \cite{types} and find that, to a good approximation, 
they are  proportional.  Thus    like the   number of  cell  types, which is  usually considered as a species complexity \cite{types}, 
$\lambda$  can also be used as an equivalent measure. We will see later, from a simple model, that $\lambda$   for a  given species is   
related to the  fraction of the total genes typically targeted   by its  miRNAs.

\begin{figure}[h]
\vspace*{.3cm}
\includegraphics[width=8.5cm]{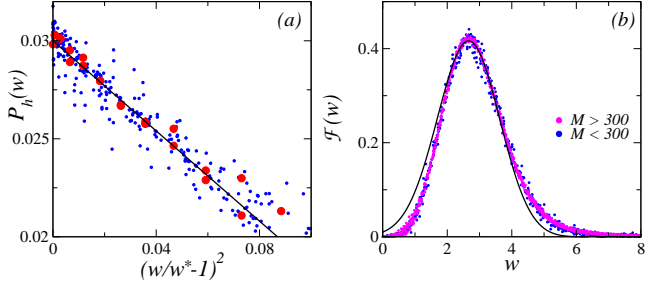}
\caption{ (a)     Weight   distribution  $P_h(w)$  for 
\textit{Homo sapiens}  (larger dots)  versus  $(w/w^*- 1)^2$ is linear   
 with slope $m=-0.115$ and $y$-intercept $c=0.030$. The small dots   denote 
 $P(w)$ of  other species  after   being collapsed to   $P_h(w).$  
(b) The scaled distribution $\Lambda P(\Lambda w)$  (dots),    
where $\Lambda=13.888\lambda$  taken from Table \ref{table:I}), 
is compared  with the universal function ${\cal F}(w),$ (eq. \eqref{eq:Fw}). 
Different colours (dots)  are used for species $M<300$ and the rest.
}
\label{fig:alpha}
\end{figure}

The  distribution of number of co-targets $P(w)$ for all the species   studied  here  
are only a scaled form of a  unique scaling  function ${\cal F}(w)$.  To   find  out this universal 
scaling  form  we make an ansatz,   
\be
{\cal F} (w)=g(w) G(w;\mu)\nonumber
\ee
where $G(w;\mu)$ is a normal distribution with mean $\mu$  and standard deviation $\sigma=1$\footnote{A Gaussian 
distribution $G^\sigma$   with  $\sigma \ne 1$  is only a  scaled form  of  $G(w;\mu),$  $i.e.$  
$G^\sigma (w;\mu) = {1 \over \sigma}G({w \over \sigma};{\mu\over \sigma})= {1 \over {\sigma \sqrt{2 \pi}}} e ^{-{{(w-\mu)^2} \over {2\sigma^2}}}.$}.
 The term $f(w)$ take care of the deviation from  normal distribution, which 
has a Taylor's   series  about $w=\mu,$ 
\bea
g(w)= g(\mu) - \alpha (w-\mu) + {\cal O}((w-\mu)^2 ), \label{Tfw}
\eea
where $\alpha \equiv -g'(\mu).$
Clearly to the $0^{th}$ order, $ {\cal F}(w)=g(\mu) G(w;\mu).$  Since  $ G(w;\mu)$ is 
already normalized, we   have $g(\mu)=1.$  To the next order in $(w-\mu),$  
\be
{\cal F}(w)= {1 \over \sqrt{2 \pi }}\left[ 1 - \alpha (w-\mu)\right]e^{-{{(w-\mu)^2} \over {2}}}.
\ee
In the present study  we will restrict ourself only to the above form of ${\cal F} (w)$ and argue that 
$P(w)$ for   different  species can be obtained   through  scaling 
\be
 P (w) = {1 \over \Lambda }  {\cal F} ({w \over \Lambda}),
\ee
where $\Lambda$  is species dependent.  
The   maximum  (or the peak) of the scaling function ${\cal F} (w)$  occurs   
at  $w=\mu^*$   where   ${\cal F}' (\mu^*) = 0.$  We find that $\mu^* = \mu - \Delta$  
where 
\bea 
 \Delta =\frac{\sqrt{1+4\alpha^2} -1}{2 \alpha} ~~ {\rm ( equivalently, } ~~ \alpha = \frac{\Delta}{1-\Delta^2} )\label{eq:Dlt}
 \eea
is  positive,   indicating  that  the added term $g(w)$  shifts peak of the 
normal distribution to  left. Thus  ${\cal F}(w)$  can be expressed  in terms of $\mu^*$ as 
\be
{\cal F}(w)={1 \over \sqrt{2 \pi }} \left[ 1 - \alpha (w-\mu^*-\Delta) \right]e^{-(w-\mu^*-\Delta)^2 / 2}.
\label{eq:Fw}
\ee
The  remaining task is to determine the parameters $\alpha$ (a measure of skewness)  and $\mu^*$ (peak position)
from  the co-target distribution  data.  However, 
the  distribution functions can not be  used directly  as  they are  scaled forms of ${\cal F}(w)$ (see eq. \eqref{eq:Fw})  
and the corresponding scale factors  are   not known. 
In fact it is enough  to determine  only the scale factor  $\hat {\Lambda}$   that  relates $P_h(w)$ with ${\cal F}(w);$
$P(w)$  of all other species which  are already collapsed onto $P_h(w)$  through $\lambda$ listed in  Table \ref{table:I}, 
can also be  collapsed onto  ${\cal F}(w)$  using  $\Lambda= \lambda \hat {\Lambda}.$  But, 
$P_h (w) = {1 \over \hat {\Lambda}} {\cal F} ({w \over \hat {\Lambda}})$  
has two unknown  parameters $\hat {\Lambda}$ and $\alpha$   which need to be determined  simultaneously. 
Note,  that  $\mu^*$  can be  calculated from knowing $\hat \Lambda$  as  $P_h(w)$ 
has  its  peak at $w^*=\hat {\Lambda}\mu^*  = 37$ (see  Table \ref{table:I}).
We  proceed  by expanding $P_h(w)$  in Taylor's series  about $w= w^*;$ to the  leading
order, 
\be
P_h(w) =  \frac{1}{\hat { \Lambda}} \left[  {\cal F} (\mu^*)  +  
  \frac{(w/w^*-  1)^2}{ \hat {\Lambda}^2 } {w^*}^2 {\cal F}^{''}( \mu^*) \right ]. 
\label{eq:Phw}
\ee
Thus  the plot of $P_h(w)$   versus  $(w/w^*- 1)^2$ 
 is expected  to be  linear  near the  peak with  slope $m= {w^*}^2 {\cal F}^{''}( \mu^*)/\Lambda^3 $  
 and $y$-intercept $c={\cal F} (\mu^*) /\Lambda.$ In fig. \ref{fig:alpha}(a) we have shown 
 this plot for  \textit{Homo sapien}  (larger dots).   The  weight distribution of all other species, 
 after collapsing  on to  $P_h(w),$  are also  plotted in the same graph
to obtain a better estimates of $m$ and $c.$  The best fitted  line,  gives  slope $m=-0.115$
and   $y$- intercept $c=0.030.$  However, from eqs. \eqref{eq:Fw} and  \eqref{eq:Phw}  we know that 
 \be
c=\frac{e^{-\Delta^2/2}}{\sqrt{2 \pi}\hat \Lambda (1-\Delta^2)}~~{\rm and}~~
m =-\frac{{w^*}^2 (1+\Delta^2)}{2\hat \Lambda^2} c~,
 \label{eq:mc}
 \ee
Evidently  $\frac{m}{c^3}$ is independent of  $\Lambda$ and
for  $\Delta \ll 1$  it can be approximated  
as $ \frac{ m}{c^3} \simeq {w^*}^2 \pi (1-3 \Delta^4/2).$ 
Using the values of $c,m$  and $w^*=37$  we have 
\bea
& &\hat \Lambda = 13.888,~ \Delta  = 0.283; \cr
&{\rm and}&  \mu^* = \frac{ w^*} {\hat \Lambda} = 2.664,~ \alpha =0.308.
\label{eq:alpha_mu}
\eea

The universal scaling function ${\cal F}(w)$  is now  specified completely.   
In fig. \ref{fig:alpha} (b)  we compare   ${\cal F}(w)$ with $\Lambda P(w \Lambda)$
using   $\Lambda=\hat {\Lambda}=13.888$  for  \textit{Homo sapien}   and   
$\Lambda =\lambda \hat{\Lambda}$  for  others, where   $\lambda$  is taken  
from  Table \ref{table:I}.   Clearly the    distributions collapse  onto 
each other   and match reasonably well  with ${\cal F} (w);$ the small  discrepancy 
observed   for $|w-\mu^*|\gg 1$ 
can  possibly  be improved  by  taking higher order  terms  of $g(w)$  in eq. \eqref{Tfw}.

\section{Model}
In the previous section we  proposed a  universal scaling  function  ${\cal F}(w)$  which   on rescaling
agrees reasonably well with the weight distribution  $P(w)$ of    miRNA   co-target networks.  It is only that 
the scale factors differ  among the  species.   ${\cal F}(w)$  has two  parameters  $\alpha$ and  $\mu^*$   which 
could be determined from the   weight distribution data.   Here we  introduce a simple microscopic 
model to understand the   origin of   the  scale  factors  $\Lambda$  and  the  constants  $\alpha$ and  
$\mu^*.$   In other words  we  would like  to understand   the dependence of $\Lambda$, $\alpha$  
and $\mu^*$  on  the number of miRNAs $M,$ the  number genes  $N,$  and the average 
number of  targets $\bar n$   of a given species.   

Let   $M$  miRNAs of a concerned species  be labeled by $i=1,2,\dots M$ and each miRNA 
$i$ targets  $n_i$ genes out of  total $N$.  Although in reality, the miRNAs  target 
specific  genes depending on  whether  it can bind to the  $3'$ UTR  of  the mRNA   (of  
the   concerned gene),  in  this model we consider  that the targets are chosen 
randomly, $i.e.$    each  miRNA $i$ target  $n_i$ genes  out of  total $N$ genes 
where $n_i$ is a stochastic variable   drawn  from  distribution  $\phi(n)$.
Since, miRNAs  bind to the $3'$ UTR  of   mRNAs,   based on the
sequence matching  and binding energies,  targets of one miRNA  is largely 
uncorrelated   with   the  targets of   the other.  Thus,   in  this random target model, 
it is  reasonable to assume that   $\phi(n)$ is a normal distribution  
with  mean  $\bar n$  and  SD  $s;$   subsequently we  denote 
$\phi(n)\equiv \phi(n; \bar n, s).$ These simple assumptions 
may  not sound very  realistic,  however we show that  it  captures the  
basic features of the  weight distribution  $P(w)$  remarkably  well.

Clearly, $n$ transcripts can be chosen   out of $N$ in $C^N_{n}$ possible ways.   
Thus, the  probability that   there are $w$ common targets among a {\it pair} 
of miRNAs, say $i=1$ and $2$, is given by 
\be
 Q_N(w,n_1,n_2) =  \frac{C^{n_1}_ w C^{N-n_1}_{n_2-w} }{  C^N_{n_2} }\;.
\label{QN}
\ee
Accordingly, the distribution of  common targets is   
\be
P(w)= \sum_{n_1,n_2=w} ^N Q_N(w, n_1,n_2) \phi(n_1; \bar n, s) \phi(n_2; \bar n, s). \label{fullPm}
\ee
   In the continuum limit, using rescaled variables 
$\nu_{1,2}=n_{1,2}/N,$ $x= w/N,$ $\bar \nu =\bar n /N,$ 
and $\sigma = s/N$   the sum   is    
converted   to  an integral
\be
P(x)= \int_{x}^1 d\nu_1 \int_{x}^1 d\nu_2 Q(x,\nu_1,\nu_2) \phi(\nu_1)\phi (\nu_2)\;, \label{eq:tildeP12}
\ee
where  all  the  functions  $P$   and $\phi$   scales as     $f(n= \nu N)  \to  f(\nu)/N.$
The functional form of  $Q(x,\nu_1,\nu_2)$,  in the large $N$  limit,  can be  obtained 
from  eq. \eqref{QN},   using  Stirling approximation,
\be Q(x, \nu_1,\nu_2)  =  \sqrt{N \over 2\pi} g(x, \nu_1,\nu_2)
e^{-N S(x, \nu_1,\nu_2)}\;\;,\ee
with $S(x, \nu_1,\nu_2)=  f(1 - \nu_1 - \nu_2 + x) + f(\nu_1 - x) + f(\nu_2 - x) +$ \\ $f(x) - f(\nu_1) - f(1 - \nu_1) - f(\nu_2) - f(1 - \nu_2),$  $f(x)=x \ln x,$\\
and 
$ g(x, \nu_1,\nu_2)= \left[{\nu_1 \nu_2(1-\nu_1)(1-\nu_2)\over x (1-\nu_1-\nu_2+x) 
(\nu_1-x)(\nu_2-x)}\right]^{1 \over 2}.
$

\begin{figure}[h]
 \includegraphics[width=8.5 cm]{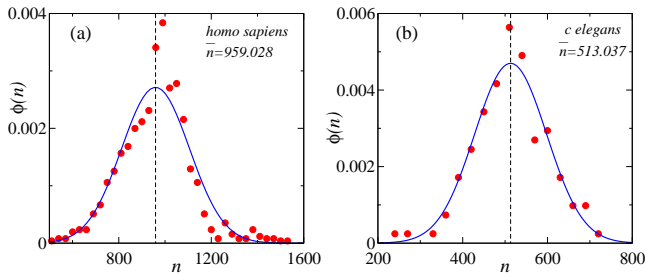}\vspace*{-.3cm}
 \caption{  Target distribution $\phi(m)$ of  miRNAs   for 
(a) \textit{Homo sapiens} and (b) \textit{C. elegans}   are fitted  to  a 
Gaussian   function (lines) with mean  and SD  ($\bar n$,  $s$) = $(959.0,147.0)$  and $(513.0,84.9)$ respectively.}
 \label{fig:phi_m_species}
\end{figure}

To proceed further, we need to specify $\phi(n),$ the distribution of the number of genes $n$ targeted by 
the miRNAs of a given  species.  We have calculated this distribution $\phi(n)$ for all 
the $22$ species;    a  plot of $\phi(n)$    is shown  in fig. \ref{fig:phi_m_species}(a) and (b)  for  
\textit{Homo sapiens} and \textit{C. elegans} respectively. The  distribution  $\phi(n)$ is too noisy as   the   sample space (total number of number 
of miRNAs   of a given species) is   too small.  Hence, it   produces  a   large error in  the estimates  
of the mean   $\bar n$ and  SD  $s,$   when fitted  to a  normal-distribution.
Moreover, we find that  the ratio of  $s$ and $\bar n$  is close to $15\%$ for 
all the species. This indicates that, for any species, about $84\%$ \footnote[1]{$\int_{\mu-\sigma}^{\mu+\sigma} G(x;\mu,\sigma)dx = Erf[1] = 0.8427$} of the number of targets  deviate at most $15\%$ from the mean $\bar n$. 
Thus,  for simplicity,   one may  assume the distribution to be $\phi(n)=\delta(n-a).$   In the following, 
we present the results  for  this choice,  as it simplifies eq. \eqref{eq:tildeP12}  substantially 
and provides a closed form expression of  ${\cal F} (w).$   There is no specific difficulty in   considering  $\phi(n)$  as a normal distribution  with  finite  width $s$;  it only scales the values of $\alpha$ and $\mu^*$  
by a $s$-dependent  factor.

\begin{figure}[h]
 \includegraphics[width=8.5cm]{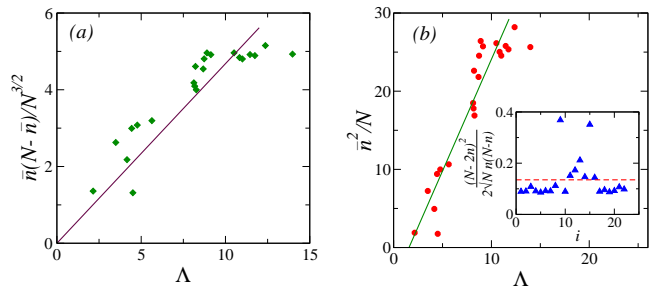}
 % validatemodel.eps: 0x0 pixel, 300dpi, 0.00x0.00 cm, bb=
 \caption{  (a) Scaling parameter 
$\Lambda=13.888\lambda$   is proportional   to  $\bar n (N - \bar n)/N^{3 \over 2},$ 
but the proportionality constant $0.468$  is different from unity  (eq. \eqref{eq:Lambda}) 
(b) $\Lambda$   versus  $\bar n^2/N$    is  linear  with slope   
$(\mu^*+\Delta)=2.54.$  Inset shows  $ \frac{(  N-  2 \bar n)^2 }{2\sqrt N  \bar n (N-  \bar n)}$  for all species and   their  average  is $\alpha=0.135$ (horizontal line).}
 \label{fig:modelvali}
\end{figure}

For $\phi(\nu)= \delta (\nu-\bar \nu)$, i.e.  when  every  miRNA  of a species target 
the same number of genes,   $P(x)= Q(x, \bar \nu, \bar \nu)$  is  significant  only  
near  $x=\bar \nu^2$ where $S(x, \bar \nu ,\bar \nu)$   has its minima.  Expanding 
both  $S(x, \bar \nu ,\bar \nu)$  and   $g(x,  \bar \nu ,\bar \nu)$ in a Taylor's  series  
about $x= \bar \nu^2$ upto the leading order we get,
 \be
P(x)=\frac{1}{\sqrt{2 \pi}\Omega}  \left[ 1-   
\frac{( 1- 2 \bar \nu)^2}{ 2 N \Omega^2}
 (x -  \bar \nu ^2)\right] e^{-{(x -\bar \nu ^2 )^2} \over {2 \Omega ^2}},  \nonumber
 \ee
where    $\Omega=  \bar  \nu (  1-  \bar  \nu)/\sqrt N.$  
 Since  $x= w/N$,  the   weight distribution $P(w)$   is related to  ${\cal F} (w)$  in 
 eq. \eqref{eq:Fw} by    the scale factor  
$\Lambda = \Omega N,$  where  
\be 
\hspace*{-.2 cm} \Lambda =\frac{\bar  n (  N-  2\bar n)} {N^{3/2}}; 
  \alpha = \frac{(  N-  2 \bar n)^2 }{2\sqrt N  \bar n (N-  \bar n)};
\mu^* +\Delta =  \frac{\bar n^2}{N \Lambda}.
\label{eq:Lambda}
\ee
In  fig. \ref{fig:modelvali}(a)  we  have shown $\bar n (  N-  2\bar n)/ N^{3/2}$ 
as a  function  of $\Lambda$  and   find that   they are proportional, but the  
proportionality constant is $0.468$ instead of unity. 
Note  that the scale-factor is a measure of the complexity of 
a species  and  now  it  can be    expressed as 
$\Lambda\simeq \bar \nu \sqrt{N}$   
because $\bar \nu,$  which  represents   the  fraction of   genes   typically targeted 
by the  miRNAs of a species,  is  usually  small (refer to Table -\ref{table:I}).  
Therefore  only the gene number is   not an  indicative of   species complexity; 
the complexity  also depends  on  `what  fraction of those genes  are  targeted by miRNAs'.

In the fig. \ref{fig:modelvali}(b)  we have also   shown  $\bar n^2/N$  as a  function
of $\Lambda$  and fit   the data   to a straight line. It follows from eq. \eqref{eq:Lambda}
that  the slope  is  $\mu^*+\Delta= 2.858.$   In the inset of this figure  we plot 
$\alpha,$ calculated using   above equation, for  all the species; the average 
value $\alpha = 0.135$  is shown as a horizontal line.  Finally  using this 
value of $\alpha$ in eq. (\ref{eq:Dlt})   we get   $\mu^* =  2.858 - \Delta =2.725.$
Clearly there is large fitting error in these estimates  of $\mu^*$ and  $\alpha$ 
and they deviate a bit from the values obtained in eq. \eqref{eq:alpha_mu}.
However, given the simplicity  of the  model where  target 
distribution  is taken as a $\delta$-function,   it  is rather   surprising   that 
the  estimates are  of the same order of magnitude as  compared to  eq. \eqref{eq:alpha_mu}.
The  difference  may be recovered  from adding a  finite  width $s$  to the 
target distribution  as  $s$  simply rescales  the  parameters   $\alpha$ and $\mu^*$
(calculations are not shown here).

\section{Conclusion}
In this article we construct miRNA co-target networks of $22$ different species,     
using the predicted miRNA targets from MicroCosm Target database \cite{microcosm}. A pair 
of miRNA are connected, only if they have at least one common target; number of 
co-targets are considered as the weight of the link. To our surprise, we find that 
the link-weight distribution of $22$ different species show an spectacular data collapse 
under scaling. Using  scaling arguments we obtain an universal scaling 
function ${\cal F}(x)$ with    two parameters:   $\mu^*$   for peak position,  
and  $\alpha$  for  skewness. The weight distribution 
functions $P(w)$s are  only  a scaled form of this function, 
$i.e., ~P(w)={\cal F}(w/\Lambda)/\Lambda;$  the scale-factors 
$\Lambda$ varies  with  species  and it may  be considered  as measure 
of complexity (number of cell types  of a  species \cite{types}).

To explain the universality, we propose a simple model where  miRNAs of 
a given species are assumed to target a fixed number of genes. This   random target model 
could provide   the correct  functional form   of ${\cal F}(x)$   and estimate  
the parameters  $\alpha$ and $\mu^*$ reasonably  well. Discrepancy   in  
these estimates  may  be substantiated by taking the  distribution  of the number of 
individual miRNA targets   as  a   Gaussian distribution with finite width. 
The model also  predicts that  the  scale-factor,  which is  a measure of 
species complexity, depends on  both of the number of genes,  and the  
fraction of genes typically targeted by  the miRNAs of that species.

It is rather surprising, why such a  simple model   captures the functional form of 
the weight distribution  of miRNA co-target network. 
Being the regulators of transcription, individual or group of miRNAs of a given species 
cooperatively target
one or more genes  for carrying out required functions. Thus, the  miRNA binding is much more 
complex than the random target model which is  quite simple  and  rudimentary. 
That it captures the weight distributions so well, rather  convey a message 
that   protein regulation  by miRNAs might have been appeared 
through some random  evolutionary process -advantageous biological functions are adopted  later
and carried forward during evolution.  Future research could reveal other 
underlying universal features of miRNA networks. 

{\it Acknowledgements :}  The authors would like to thank  Prof. Ayse  Erzan  for 
helpful discussions.

\end{document}